\documentclass[english]{article}
\usepackage[T1]{fontenc}
\usepackage[latin9]{inputenc}
\usepackage{amstext}
\usepackage{amssymb}
\usepackage{graphicx}
\usepackage{esint}

\makeatletter
\newcommand{\lyxaddress}[1]{
\par {\raggedright #1
\vspace{1.4em}
\noindent\par}
}

\usepackage{a4wide}

\makeatother

\usepackage{babel}
\begin{document}

\title{The Aharonov Casher Effect: The Case of $g\neq2$}

\author{Niv Cohen\thanks{Cohen.niv@gmail.com} $\;$and Oded Kenneth}

\maketitle

\lyxaddress{%
\begin{minipage}[t]{1\columnwidth}%
Department of Physics \\
Technion, Israel Institute of Technology, \\
Haifa 32000, Israel \\%
\end{minipage}}
\begin{abstract}
The Aharonov Casher effect predicts the existence{\normalsize{} in
two dimensions} of {\normalsize{}$\left\lceil \frac{\mbox{\ensuremath{\Phi}}}{2\pi}\right\rceil -1$
bounded zero modes associated with }a magnetic flux $\Phi${\normalsize{}.}
Aharonov and Casher discussed the case of gyromagnetic factor equals
2, we will discuss the general case of any gyromagnetic factor. As
a simple model, we study the case where the magnetic field lies in
a thin annulus. First we examine the wavefunctions of the zero-energy
bounded states, predicted by the Aharonov Casher Effect for electrons
with gyromagnetic ratio equal 2. We then calculate the wave function
and energies for a gyromagnetic ratio $g\neq$ 2. We give the dependence
of the bound states energies on g and the angular momentum. Finally,
we provide an order of magnitude estimations for the binding energies.
\end{abstract}

\section{Aharonov Casher zero modes }

In this section we will find explicitly the Aharonov Casher zero modes
for a magnetic field which lives inside a thin annulus. First we write
the Pauli equation for an electron in a plane. We take the magnetic
field to be in the direction normal to the plane and the electron's
spin to be along the field. Taking units where $\hbar=c=2m_{e}=1$
we get:

\begin{equation}
\left(\left(-i\nabla-A\left(r\right)e\right)^{2}+\frac{g}{2}B\left(r\right)e\right)\psi=0
\end{equation}
From now on, we will measure magnetic flux in units of $\frac{1}{e}$.
In our case, the entire magnetic flux lives on a infinitesimal thin
circle of radius $R$:

\[
B\left(r\right)=\delta\left(R-r\right)\cdot\frac{\Phi}{2\pi R}
\]
We will work in the Coulomb gauge:

\begin{eqnarray*}
r<R & :\  & A=0\\
r>R & :\  & A\left(r\right)=(\Phi/2\pi r)\hat{\theta}
\end{eqnarray*}
and in angular coordinates:

\begin{eqnarray}
r<R: &  & \left(-\frac{1}{r}\partial_{r}r\partial_{r}+\bigl(-i\partial_{\theta}/r\bigr){}^{2}\right)\psi_{in}=0
\end{eqnarray}
\begin{eqnarray}
r>R: &  & \left(-\frac{1}{r}\partial_{r}r\partial_{r}+\left(-i\partial_{\theta}/r-\Phi/2\pi r\right){}^{2}\right)\psi_{out}=0
\end{eqnarray}
The Pauli equation (1) around at $r=R$ gives the jump conditions
between $\psi_{in}$ and $\psi_{out}$. In a small enough neighborhood
of $r=R$ we can keep only second derivatives in $r$ and the Stern-Gerlach
term, while neglecting all the other terms with respect to them, to
get: $(-\partial_{r}^{2}-\frac{g}{2}B)\psi=0$. Integrating $\partial_{r}^{2}$
in a small neighborhood around $r=R$, we get the jump condition of
$\psi$ between the two domains, $r<R$ and $r>R$ :

\begin{equation}
(\partial_{r}\psi_{out}-\partial_{r}\psi_{in})|_{\text{r=R}}=-\frac{g}{2}\frac{\Phi}{2\pi R}\psi
\end{equation}
In order to find the wavefunction in the two domains, we will use
the separation of variables: $\psi=f\left(r\right)e^{i\theta m}$
($m\in\mathbb{Z}$), and get:\footnote{For $m=0,$ the second term of $\psi_{in}$ is $D_{0}\ln\left(r\right)$,
and is still not continuous at 0. A similar correction should be noted
for the solutions of $\psi_{out}$, where the solution containing
a logarithmic function is rejected for not being normalizable.}. 

\[
\psi_{in}=C_{0}r^{m}e^{im\theta}+D_{0}r^{-m}e^{im\theta}
\]

\[
\psi_{out}=E_{0}r^{m-\frac{\mbox{\ensuremath{\Phi}}}{2\pi}}e^{im\theta}+F_{0}r^{-(m-\frac{\mbox{\ensuremath{\Phi}}}{2\pi})}e^{im\theta}
\]
From the restrictions of continuity, normalization and matching of
the jump condition we must take $D_{0}=F_{0}=0$ and $m\geq0$. The
jump condition gives us $C_{0}=E_{0}$, which also give us continuity.
Note that the existence of the Aharonov and Casher zero mode seems
to be very sensitive to the jump condition, determined by the value
of $g$. The remaining wavefunction is normalizable if, and only if,
$\psi_{out}$ is normalizable. Therefore we require:

\begin{eqnarray*}
\intop_{R}^{\infty}|\psi|^{2}2\pi rdr & = & \intop_{R}^{\infty}\left(r^{(m-\frac{\mbox{\ensuremath{\Phi}}}{2\pi})}\right){}^{2}r2\pi dr<\infty
\end{eqnarray*}
or $m<\frac{\mbox{\ensuremath{\Phi}}}{2\pi}-1$. This gives us the
$\left\lceil \frac{\mbox{\ensuremath{\Phi}}}{2\pi}\right\rceil -1$
different solutions predicted by Aharonov and Casher.

\section{Bounded states with correction to the gyromagnetic factor}

In this section we will find the wavefunctions and binding energies
for the case of $g>2$ (the case of $g<2$ does not yield a normalizable
wavefunctions). Using the same separation of variables as before:
$\psi=f\left(r\right)e^{i\theta m}$, we get:

\[
r<R:\left(-\frac{1}{r}\partial_{r}r\partial_{r}+m^{2}/r^{2}\right)f\left(r\right){}_{in}=Ef\left(r\right){}_{in}
\]

\[
r>R:\left(-\frac{1}{r}\partial_{r}r\partial_{r}+\left(m-\left(\Phi/2\pi\right)\right){}^{2}/r^{2}\right)f\left(r\right){}_{out}=Ef\left(r\right){}_{out}
\]

Looking at $E<0$, we express $f\left(r\right)$ as a linear combination
of the modified Bessel functions of the first and second kind $\left(I_{n,}K_{n}\right)$: 

\begin{eqnarray*}
f\left(r\right){}_{in} & = & C_{1}I_{m}\left(\sqrt{-E}r\right)+C_{2}K_{m}\left(\sqrt{-E}r\right)\\
f(r)_{out} & = & C_{3}I_{\left(m-\left(\Phi/2\pi\right)\right)}\left(\sqrt{-E}r\right)+C_{4}K_{\left(m-\left(\Phi/2\pi\right)\right)}\left(\sqrt{-E}r\right)
\end{eqnarray*}
For normalization and continuity at 0, we must choose $C_{2}=C_{3}=0$.
From continuity at $r=R$ we get:

\[
\frac{C_{1}}{C_{4}}=\frac{K_{\left(m-\left(\Phi/2\pi\right)\right)}\left(\sqrt{-E}R\right)}{I_{m}\left(\sqrt{-E}R\right)}
\]
The binding energy$E$ depends on the following parameters: $n,\Phi,g$
. Naturally, we are interested in the function $E(g)$, for a value
of g slightly above 2. It is technically simpler to examine $g\left(m,\Phi,E\right)$,
which we get explicitly from jump condition (4):
\[
g(E)=-(\partial_{r}\psi_{out}-\partial_{r}\psi_{in})|_{\text{r=R}}\frac{4\pi R}{\Phi}/\psi(R)
\]

or:

\[
g(E)=\left(\frac{\partial_{r}I_{m}\left(\sqrt{-E}r\right)|_{\text{r=R}}}{I_{m}\left(\sqrt{-E}R\right)}-\frac{\partial_{r}K_{\left(m-\left(\Phi/2\pi\right)\right)}\left(\sqrt{-E}r\right)|_{\text{r=R}}}{K_{\left(m-\left(\Phi/2\pi\right)\right)}\left(\sqrt{-E}R\right)}\right)\frac{4\pi R}{\Phi}
\]
For $m=0$, Expanding to first order in $E$ we get:

\[
g(E)=2-\frac{R^{2}E}{2(\frac{\Phi}{2\pi}-1)}+O(E^{\frac{\Phi}{2\pi}})
\]

or:
\[
E(g)=-\frac{(g-2)\cdot(\frac{\Phi}{2\pi}-1)}{R^{2}}+O\left(\left(g-2\right)^{\frac{\Phi}{2\pi}}\right)
\]

As an example, we can plot the electron\textquoteright s energy as
function of the flux (noted as phi in the graph), for the gyromagnetic
ratio of an isolated electron given by QED and no orbital angular
momentum. We plot it for $R=1/\alpha^{2}$ (or $R=\frac{\hbar}{2m_{e}c\alpha^{2}}=3.626\cdot10^{-9}meter$)
\footnote{See section 3.1 for the reason we chose this radius}. $\Phi$
is shown in units of $\frac{\Phi}{2\pi}$ (or $\phi_{unit}=\frac{\text{h}}{e}=4.135\cdot10^{-15}Wb$
) and $E$ is of order of magnitude of $m_{e}c^{2}\alpha^{5}$ and
shown in $eV$.

\includegraphics[scale=0.6]{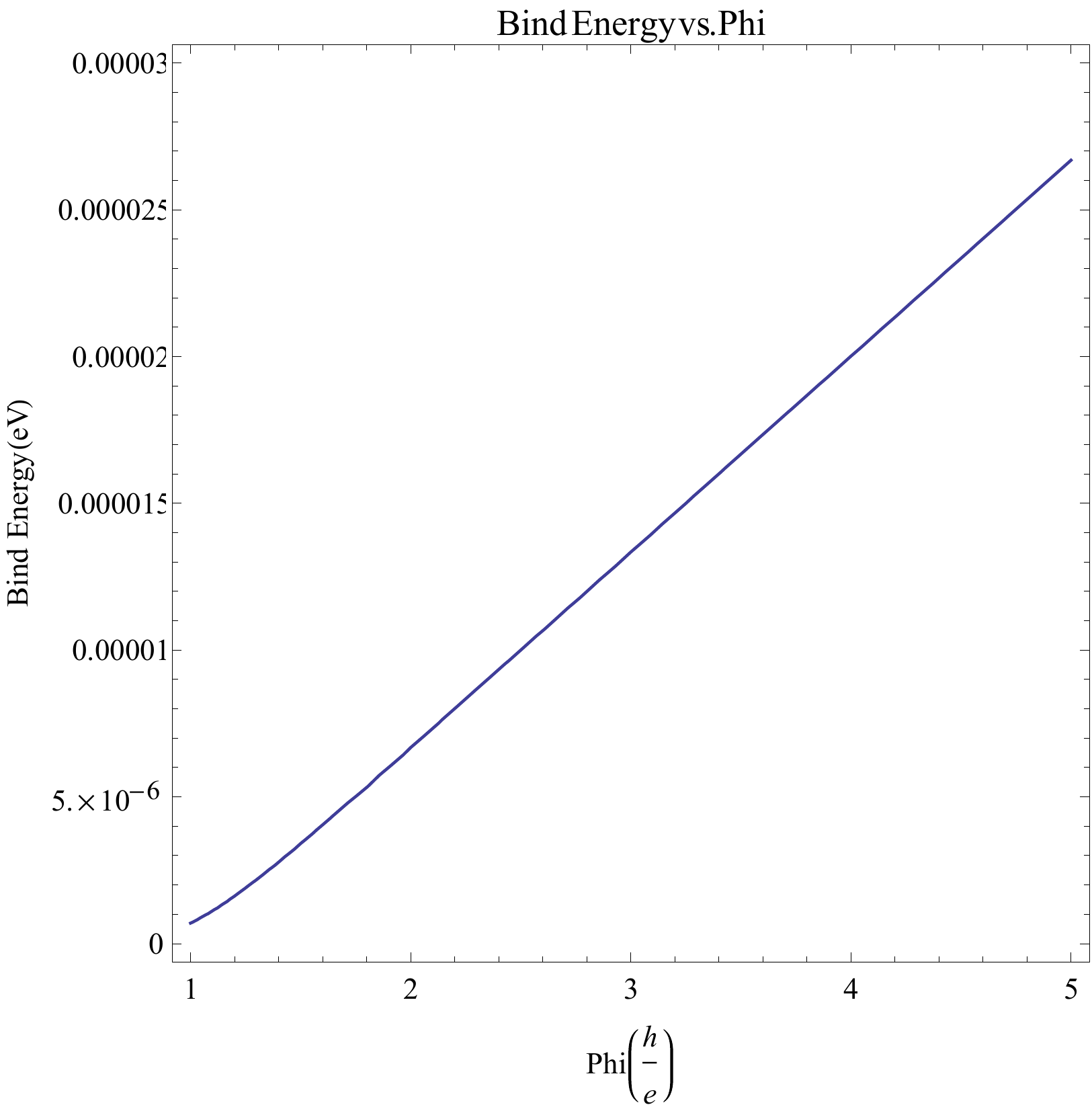}

We can see that the binding energy near $\frac{\Phi}{2\pi}=1$ is
indeed of an order of magnitude less than the first order approximation,
and that the slope is consistent with our approximation as well. 

Another interesting relation is the binding energy as a function of
the gyromagnetic ratio. For example, we plot it for a fixed $\frac{\Phi}{2\pi}=2$:

\includegraphics[scale=0.6]{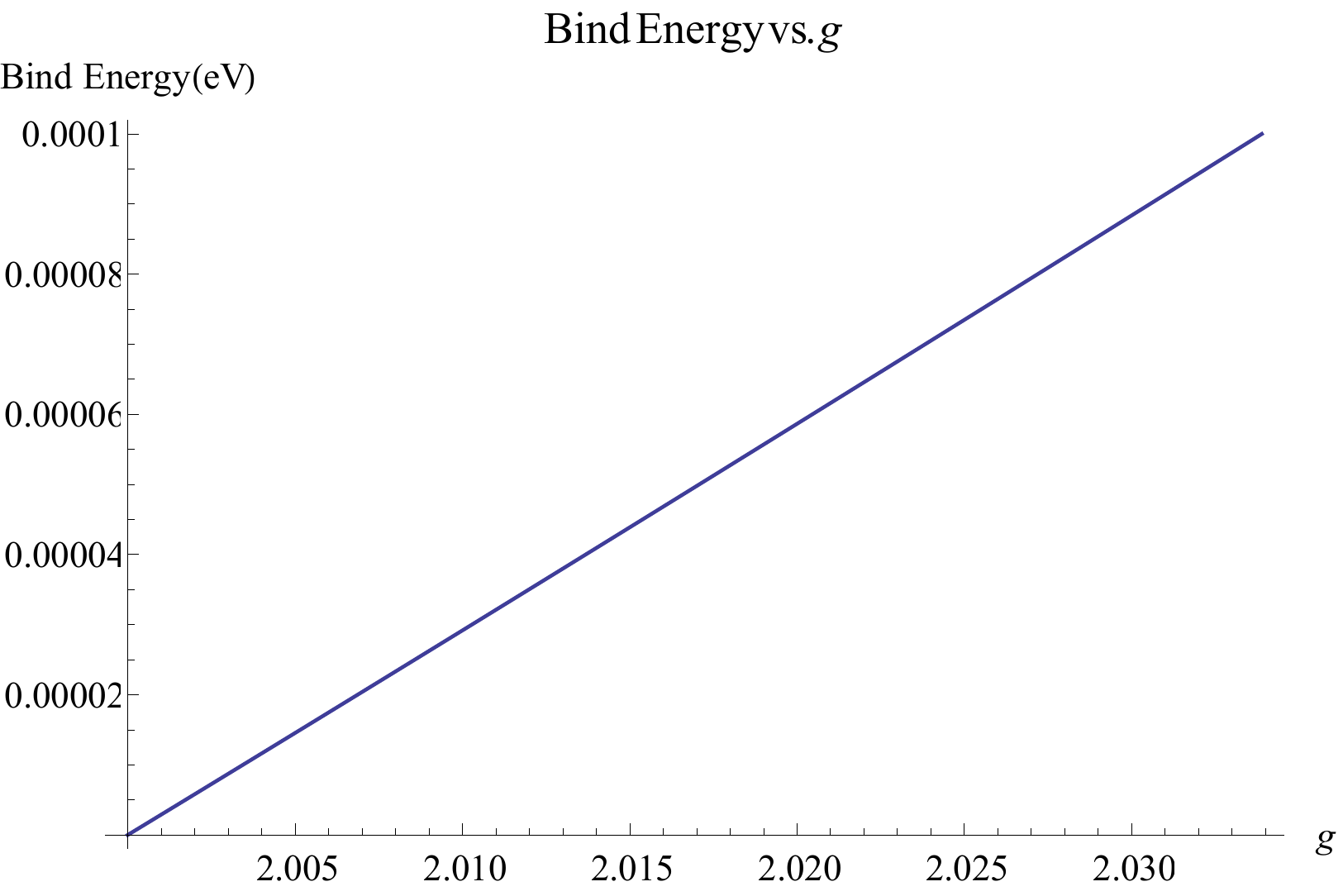}

We should also note two other interesting properties of those states:
First, the critical flux for the emergence of a new bound state is
somewhat lower with respect to the value given for $g=2$. Second,
the decay rate of the states far from the magnetic flux has a typical
radius, in contrast with the $g=2$ power law case.

\section{Order of magnitude}

In this section we will work in m.k.s units.

\subsection{Estimation of the source of a unit magnetic flux:}

A flux that is equal $2\pi$ in the units we used previously, is equal
$\Phi_{unit}=2\pi h/e$ in the m.k.s. units. One possible way to produce
such a flux is to use a collection of magnetic dipoles originating
from an electron's orbital angular momentum or spin. We will use the
semi-classical Bohr model in order to estimate how many such atoms
will be needed in order to get a one unit of quantum flux. Our model
is a current loop, with a current that matches a single electron with
the velocity of$\alpha c$, where $\alpha$ is the fine structure
constant. This gives a current of $I=\frac{e\alpha c}{2\pi a_{0}}$
in a loop of radius $a_{b}$ (the Bohr radius). By the Biot Savart
law we get:

\[
B_{_{max}}=\frac{e\alpha c}{2\pi a_{b}}\frac{\mu_{0}}{r}
\]
For small radii around the current loop, the contribution to the flux
from one side of the loop cancels the contribution from the other
side. In a distance similar to the Bohr radius from the loop, there
is no longer a similarity between the magnetic field in the two sides
of the current loop so $B_{center}\sim\frac{e\alpha c}{2\pi a_{b}}\frac{\mu_{0}}{a_{b}}$.
This typical value lasts for an area of about $\pi a_{b}^{2}$ which
gives us:

\[
\Phi\sim B_{center}A=\frac{e\alpha}{\epsilon_{0}c}
\]

or:

\[
\Phi/\Phi_{q}\sim\alpha^{2}
\]
\\
This means, we will need a magnetic tip with a cross section of order
of magnitude of $1/\alpha^{2}$ dipoles in order to produce one unit
of quantum flux at the end of the tip.

A second option to produce such a flux, is to pass a strong magnetic
flux through a superconductor of the second kind, such that the penetrations
of the magnetic flux will be dense enough for spotting the effect.

\subsection{Estimation of  the binding energy}

As we saw in section 2, the bind energy is of the order of magnitude
of:

\[
E\sim2m_{e}c^{2}(g-2)\cdot(\frac{\Phi}{\Phi_{q}}-1)\frac{\left(\frac{\hbar}{2m_{e}c}\right)^{2}}{R^{2}}
\]
For an electron with the QED vacuum correction to the gyromagnetic
factor and a magnetic field in a thin annulus of radius r, we get:

\[
E\sim\alpha\frac{h^{2}}{m_{e}R^{2}}\left(\frac{\Phi}{\Phi_{q}}-1\right)
\]
In comparison to the hydrogen atom bind energy $E_{h}\sim\frac{h^{2}}{a_{b}^{2}m_{e}}$,
we get:

\[
E/E_{h}\sim\alpha\frac{a_{b}^{2}}{R^{2}}\left(\frac{\Phi}{\Phi_{q}}-1\right)
\]
Substituting a flux of a size of a few unit fluxes, according to the
estimation from section (3.1), we get:

\[
E/E_{h}\sim\alpha^{3}
\]

\section*{Acknowledgment}

I am very thankful to Prof. Yosi Avron for his guidance and useful
discussions during this research. The work of OK is supported by the
ISF.

\end{document}